\documentclass[aps,floats,subeqn,showpacs,amssymb]{revtex4}
\usepackage{epsfig}
\usepackage{epstopdf}
\usepackage{float} 
\usepackage{caption}
\usepackage[hang,nooneline,large]{subfigure}
\DeclareGraphicsExtensions{.pdf,.eps,.png,.jpg,.mps} 
\date{\today}
\usepackage{amsmath}
\usepackage{amssymb}

\newcommand{\be}{\begin{equation}}
\newcommand{\ee}{\end{equation}}
\newcommand{\bea}{\begin{eqnarray}}
\newcommand{\eea}{\end{eqnarray}}
\newcommand{\bml}{\begin{mathletters}}
\newcommand{\eml}{\end{mathletters}}

\begin{document}

\tighten

\draft




\title{Gauss-Bonnet boson stars}
\renewcommand{\thefootnote}{\fnsymbol{footnote}}

\author{Betti Hartmann $^{(a)}$ }
\email{b.hartmann@jacobs-university.de}
\author{J\"urgen Riedel $^{(a), (b)}$}
\email{jriedel@thescienceinstitute.com}
\author{Raluca Suciu $^{(a)}$}
\email{r.suciu@jacobs-university.de}

\affiliation{
$(a)$ School of Engineering and Science, Jacobs University Bremen, 28759 Bremen, Germany\\
$(b)$ Institut f\"ur Physik, Universit\"at Oldenburg, 26111 Oldenburg, Germany}

\date{\today}
\setlength{\footnotesep}{0.5\footnotesep}

\begin{abstract}
We construct boson stars in (4+1)-dimensional Gauss-Bonnet gravity. We study the properties of the solutions
in dependence on the coupling constants and investigate these in detail. While the ``thick wall''
limit is independent of the value of the Gauss-Bonnet coupling, we find that the spiraling
behaviour characteristic for boson stars in standard Einstein gravity disappears for large enough values of the 
Gauss-Bonnet coupling. Our results show that in this case the scalar field can {\it not} have arbitrarily high values
at the center of the boson star and that it is hence impossible to reach 
the ``thin wall'' limit. Moreover, for large enough Gauss-Bonnet coupling we find a {\it unique} relation between the 
mass and the radius (qualitatively similar to 
those of neutron stars) 
which is not present in the Einstein gravity limit. 

\end{abstract}

 \maketitle
\section{Introduction}
Solitons are localized and finite energy solutions that are in general non-singular and stable and
as such can be seen as models for particles. 
In the context of relativistic field theories, solitons are typically
classified into two distinct groups, namely ``topological solitons'' and ``non-topological solitons''.
Topological solitons \cite{ms} require degenerate vacuum states. 
The topological character of the field is represented by an integer which is called the topological charge. 
Topological solitons result (in most cases) from a spontaneous symmetry breaking.
Non-topological solitons, on the other hand, arise in field theories with unbroken continuous symmetries. 
Examples of non-topological solitons are $Q$-balls \cite{fls,lp,coleman}.
These are made out of scalar fields prevented from collapse by Heisenberg's uncertainty principle
and repulsive self--interaction.
They carry a non-vanishing Noether charge $Q$ that is globally conserved due to the global U(1) symmetry
of the model. 
$Q$ can e.g. be interpreted as particle number \cite{fls}. 
As such, $Q$-balls have been constructed in $(3+1)$-dimensional models with 
non-renormalizable scalar field potential \cite{vw,kk1,kk2}, but also appear in supersymmetric 
extensions of the Standard Model \cite{kusenko,cr,ct}.
These supersymmetric $Q$-balls have been considered as possible candidates for baryonic dark matter 
\cite{dm} and their astrophysical implications have been discussed \cite{implications}.    

The self-gravitating counterparts of $Q$-balls, so-called ``boson stars'' have also been discussed extensively
\cite{kaup,misch,flp,jetzler,new1,new2,Liddle:1993ha}. Since the discovery of an elementary scalar particle in nature \cite{cern_lhc}
it is by itself interesting to study these type of objects. Even if observations would
exclude them to exist they
are described by relatively simple equations and could hence act as simple toy models for a
wide range of objects such as
particles, compact stars, e.g. neutron stars and even centers of 
galaxies \cite{schunck_liddle}. 
     
In this paper we are interested in boson stars in the context of $(4+1)$-dimensional Gauss-Bonnet theory which
appears naturally in the low energy effective action of quantum gravity models 
 \cite{zwiebach}. 
An important property of Gauss-Bonnet gravity is that its spectrum 
does not include new propagating degrees of freedom besides gravitation. 
In $(3+1)$ dimensions, the Gauss-Bonnet term is a total derivative so it only 
contributes to the field equations when it is coupled to a dilaton field. Here, we are only interested in the
effect of Gauss-Bonnet gravity and will hence study these objects in the minimal number of dimensions
in which the term does not become a total derivative. In \cite{Pani:2011xm} compact stars consisting of
a perfect fluid have been studied
in modified models of gravity, including Einstein-Gauss-Bonnet-dilaton
gravity in $(3+1)$ dimensions. Let us remark that boson stars are not made out of a perfect fluid
since the diagonal spatial components of the energy-momentum tensor are not all equal. 

The interest in studying boson stars in higher dimensions is also supported by several other arguments. 
First of all, higher dimensions appear in attempts to find a quantum description of gravity as well
as in unified models. Examples are Kaluza-Klein theories and String Theory. 
For black holes it became clear that many of their properties in $(3+1)$ dimensions do not extend to 
higher dimensions. It is therefore natural to consider other 
localized objects in higher dimensions such as boson stars and see which influence the number of dimensions
has on their properties. 
$Q$-balls and boson star solutions of the full system of coupled non-linear equations 
in $(4+1)$-dimensional asymptotically flat
space-time have been investigated in \cite{hartmann_kleihaus_kunz_list,hartmann_riedel2,hartmann_riedel1} and
it was indeed found that the behaviour of the solutions depends crucially on the number of spatial dimensions.

Also, if we add a negative cosmological 
constant we obtain boson stars in $(4+1)$-dimensional Anti-de Sitter (AdS) space-time and according to the 
AdS/CFT correspondence \cite{ggdual,adscft} this theory would correspond to a $(3+1)$-dimensional strongly coupled conformal field theory
on the boundary of AdS. As such boson stars have been suggested to be the dual of condensates of glueballs 
\cite{horowitz} and their properties in $(d+1)$ dimensions have been investigated \cite{hartmann_riedel2,
hartmann_riedel1}. 

Finally, it appears technically simpler to rotate objects in higher dimensions. 
In the case of $(3+1)$ dimensions rotating boson stars are axially symmetric \cite{vw,kk1,kk2}, so we 
need to solve partial differential equations, while in $(4+1)$ dimensions we have two
planes of rotation. If we choose the two angular momenta associated with the 
two orthogonal planes equal to each other the symmetry of the solutions can be enhanced and ``only'' 
ordinary differential equations need to be solved \cite{hartmann_kleihaus_kunz_list}.
     
Our paper is organized as follows: in Section II we give the model, equations of motion as well as the
definition of mass, charge and radius of our solutions. In Section III we discuss our numerical results, while
in Section IV we conclude. 

\section{The model and the equations}

In this paper we study boson stars in 5-dimensional Gauss-Bonnet gravity. 
The action reads~:
\begin{equation}
S= \int d^5 x \sqrt{-g} \left(R   + 
\alpha\left(R^{MNKL} R_{MNKL} - 4 R^{MN} R_{MN} + R^2\right) + 16\pi G_5 {\cal L}_{\rm matter}\right) \ ,
\end{equation}
where $\alpha$  is the Gauss--Bonnet coupling and
$\alpha=0$ corresponds to Einstein gravity. $G_{5}$ is Newton's constant in 5 dimensions
which is connected to the 5-dimensional Planck mass $M_{\rm pl,5}$ by $G_5={M^3_{\rm pl,5}}$.
${\cal L}_{\rm matter}$ denotes the matter Lagrangian of the complex valued scalar field $\psi$ which reads~:
\begin{equation}
{\cal L}_{\rm matter}= -\left(\partial_{\mu}\psi\right)^* \partial^{\mu} \psi - U(\psi)
\end{equation}
where the scalar potential reads
\begin{equation}
U(\psi)=m^2\eta_{\rm susy}^2 \left(1-\exp\left(-\frac{\vert\psi\vert^2}{\eta_{\rm susy}^2}\right)\right)  \ .
\end{equation}
This potential can be developed into a series as follows
\begin{equation}
\label{potentialphi6}
U(\psi)=m^2 \vert\psi\vert ^2-\frac{m^2 \vert\psi\vert ^4}{2 \eta_{\rm susy} ^2}
+\frac{m^2 \vert\psi\vert^6}{6 \eta_{\rm susy}
^4}+O\left(\vert\psi\vert ^8\right)   \ . \end{equation}
In the following, we will use the potential up to 6th order in $\psi$. We have checked that the 
results are qualitatively similar, however, it turns out that the numerics at critical points is easier
with a $\psi^6$ potential. 

The gravity equations are obtained from the variation of the
action with respect to the metric fields and read
\begin{equation}
 G_{MN} +  \frac{\alpha}{2} H_{MN}=8\pi G_5 T_{MN} \ ,  M,N=0,1,2,3,4 \ ,
\end{equation}
where $H_{MN}$ is given by
\begin{eqnarray}
 H_{MN}&=& 2\left(R_{M ABC}R_{N}^{ABC} - 2 R_{M A N B}R^{AB} - 2 R_{M A}R^{A}_{N} + 
R R_{MN}\right)  \nonumber \\
&-& \frac{1}{2} g_{MN} \left(R^2 - 4 R_{AB}R^{AB} + R_{ABCD} R^{ABCD}\right) \ , \ \ A,B,C=0,1,2,3,4 \ , 
\end{eqnarray}
and $T_{MN}$ is the energy-momentum tensor
\begin{eqnarray}
\label{em}
T_{MN}&=& g_{MN} {\cal L} - 2\frac{\partial {\cal L}}{\partial g^{MN}}\nonumber\\
&=& -g_{MN} \left[\frac{1}{2} g^{KL} 
\left(\partial_{K} \psi^* \partial_{L} \psi +
\partial_{L} \psi^* \partial_{K} \psi\right) + U(\psi)\right] +
\partial_{M} \psi^* \partial_{N} \psi + \partial_{N}\psi^* \partial_{M} \psi  \ .
\end{eqnarray}

The variation of the action with respect to the matter field leads to 
the Klein-Gordon equation which reads
\begin{equation}
\label{KG}
 \left(\square - \frac{\partial U}{\partial \vert\psi\vert^2} \right)\psi=0 \ \   \ \ .
\end{equation}
The matter Lagrangian ${\cal L}_{matter}$  is invariant under the global U(1) transformation
\begin{equation}
 \psi \rightarrow \psi e^{i\chi} \ \ \  .
\end{equation}
As such the locally conserved Noether
current $j^{M}$, $M=0,1,..,4$ associated to this symmetry is given by
\begin{equation}
j^{M}
 = -\frac{i}{2} \left(\psi^* \partial^{M} \psi - \psi \partial^{M} \psi^*\right) \  \ {\rm with} \ \ \
j^{M}_{; M}=0  \ .
\end{equation}
The globally conserved Noether charge $Q$ of the system then reads
\begin{equation}
 Q= -\int d^{4} x \sqrt{-g} j^0  \  .
\end{equation}

\subsection{Ansatz and equations of motion}

We choose the following Ansatz for the metric~:
\begin{equation}
ds^2 = - N(r) A^2(r) dt^2 + \frac{1}{N(r)} dr^2 + r^2 \left(d\theta^2 + \sin^2\theta d\varphi^2 + 
\sin^2\theta \sin^2\varphi d\chi^2\right)
\end{equation}
where $N$ and $A$ are functions of $r$ only and $\theta,\varphi,\chi$ are angular coordinates. We further choose
\begin{equation}
 N(r)=1-\frac{2n(r)}{r^{2}}  \  , 
\end{equation}
such that $n(\infty)$ will determine the gravitational mass of the solution. 
For the scalar field we choose
\begin{equation}
 \psi(r,t)=f(r) e^{i\omega t}  \ ,
\end{equation}
where $\omega$ is the internal frequency. In general, boson star solutions
exist only in a limited parameter range of $\omega$. This parameter range depends on the choice of 
potential. For our potential (\ref{potentialphi6}) we have $\omega \in [0:1]$ \cite{hartmann_riedel1}.

We introduce the following rescalings
\begin{equation}
\label{rescale}
 r\rightarrow \frac{r}{m} \ \ , \ \ \omega \rightarrow m\omega \ \ , \ \  
 \psi\rightarrow \eta_{\rm susy} \psi \ \ , \ \ n\rightarrow n/m^{2}  \ \ , \ \ \alpha\rightarrow \alpha/\sqrt{m}
\end{equation}
and find that the equations depend only on the dimensionless coupling constants $\alpha$ and
\begin{equation}
 \kappa=8\pi G_{5}\eta_{\rm susy}^2   \ .
\end{equation} 

The equations of motion are
\begin{equation}
\label{eq_a}
   A'=\frac{2 \kappa  r^3 \left(A^2 N^2 f'^2+\omega ^2 f
   ^2\right)}{3 A N^2 \left(2 \alpha -2 \alpha  N+r^2\right)}   \ ,
\end{equation}
\begin{equation}
\label{eq_n}
\begin{split}
   N'(r)=2 r\left(\frac{1-N}{r^2+2\alpha(1-N)}\right)
   -\frac{2}{3}\frac{\kappa r^3}{N A^2}\frac{(1-e^{-f^2}) N A^2+\omega^2 
f^2+N^2 A^2 f'^2}{r^2+2\alpha(1-N)}   \ ,
  \end{split}
\end{equation}
for the metric functions and
\begin{equation}
\label{phi_eq}
 \left(r^{3} A N f'\right)'= r^{3} A 
\left(\frac{1}{2} \frac{\partial U}{\partial f} -\frac{\omega^2f}{N A^2}\right)  \ 
\end{equation}
for the matter field function. Here and in the following the prime will denote the derivative with respect to $r$.

These equations have to be solved numerically subject to appropriate boundary conditions.
We want to construct globally regular solutions with finite energy. At the origin we hence require
\begin{equation}
\label{bc1}
f'(0)=0 \ \ , \ \ \ n(0)=0 \ .
\end{equation}
Moreover, the scalar field function falls of exponentially with
\begin{equation}
\label{bc2a}
 f(r>>1)\sim \frac{1}{r^{\frac{3}{2}}} \exp\left(-\sqrt{1-\omega^2}r\right) + ...
\end{equation}
and we hence require $f(\infty)=0$, while we choose
$A(\infty)=1$ (any other choice would just result in a rescaling of the time coordinate).

\subsection{Mass, charge and radius}
As in most cases considered so far the scalar field function falls off exponentially in our case.
Hence, there are different possibilities to define the ``radius'' of our boson star solutions.
Let us remark that models with a V-shaped potential have been considered  \cite{Arodz:2008jk,Arodz:2008nm} 
that possess compact boson stars with a well-defined outer radius (very similar to those of ``standard stars'')\cite{Kleihaus:2009kr,Kleihaus:2010ep,Hartmann:2012da}. 
Here we follow \cite{radu} and define the radius of the boson star as an averaged radial coordinate
\begin{equation}
\label{radius}
 R= \frac{2 \pi^2}{Q} \int\limits_0^{\infty} {\rm d} r \ r^{4} \frac{\omega f^2}{AN} \ .
\end{equation}

Furthermore, the explicit expression for the Noether charge reads
\begin{equation}
\label{charge_ex}
 Q=2 \pi^2 \int\limits_0^{\infty} {\rm d} r \ r^{3} \frac{\omega f^2}{AN} \ .
\end{equation}

For $\kappa=0$ we have $A\equiv 1$ and $n\equiv 0$. Then the mass 
$M$ corresponds to the integral of the energy density $\epsilon=-T^0_0$ 
and reads
\begin{equation}
\label{mass}
 M=2 \pi^{2} \int\limits_0^{\infty} {\rm d} r \ r^{3} \  \left(N \phi'^2 + 
\frac{\omega^2 \phi^2}{N} + U(\phi)\right)  \ .
\end{equation}
For $\kappa\neq 0$ the same procedure would lead to the following integral
\begin{equation}
 M_I=2 \pi^{2} \int\limits_0^{\infty} {\rm d} r \ r^{3} A \  \left(N \phi'^2 + 
\frac{\omega^2 \phi^2}{N} + U(\phi)\right)  \ .
\end{equation}
In the following, we will refer to the mass as to the ``inertial mass''.
Using the equation of motion (\ref{eq_n}) this reads 
\begin{equation}
M_I=\frac{6\pi^2}{\kappa} \int\limits_0^{\infty} dr A n'  \ . 
\end{equation}
Clearly, if $A\equiv 1$ the inertial mass would be given in terms of $n(\infty)$. However, in general, we will
have $A\neq 0$. We hence define the {\it gravitational mass} to be given by the asymptotic behaviour of the metric function
$M_G\sim n(r \rightarrow \infty)/\kappa$. As shown in \cite{amsel_gorbons} this procedure can also be applied in the
case of Gauss-Bonnet gravity.

\subsection{Corresponding black hole solutions}
The equations of motion (\ref{eq_a}) - (\ref{phi_eq}) possess black hole solutions for $f\equiv 0$ which implies
$A\equiv 1$. The metric function $N(r)$ is then given by:
\begin{equation}
 N(r)=1 + \frac{r^2}{2\alpha} \left( 1 - \sqrt{1+ \frac{8\alpha M}{r^4}}\right)   \ ,
\end{equation}
where $M$ is an integration constant that corresponds to the mass of the solution. This is the Boulware-Deser
solution \cite{boulware_deser}. Note that for $\alpha \rightarrow 0$
this becomes $N(r)=1-2M/r^2$, which is the Tangherlini-Schwarzschild solution in $(4+1)$ dimensions \cite{tan}. This solution
has an event horizon at 
\begin{equation}
 r_h=\sqrt{2M-\alpha}  \ .
\end{equation}
The gravitational mass of these black hole solutions is given by $M_G=\lim_{r\rightarrow\infty} n(r)/\kappa=M/\kappa$. 

\begin{figure}[h!]
\begin{center}
\subfigure[\ $\kappa=0.02$]
{\label{fig_charge1}\includegraphics[width=5.5cm,angle=270]{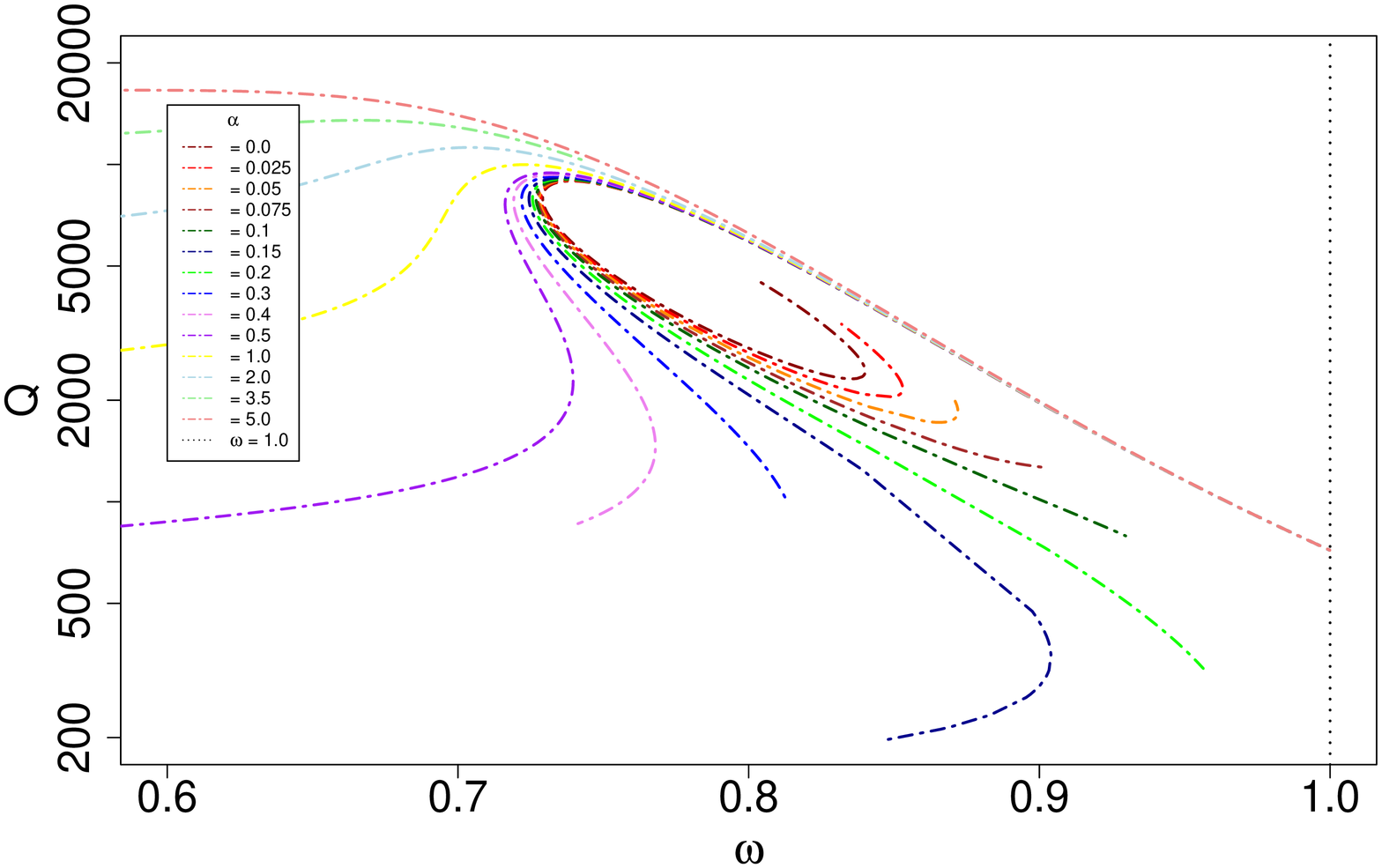}}
\subfigure[\ $\kappa=0.05$]
{\label{fig_charge2}\includegraphics[width=5.5cm,angle=270]{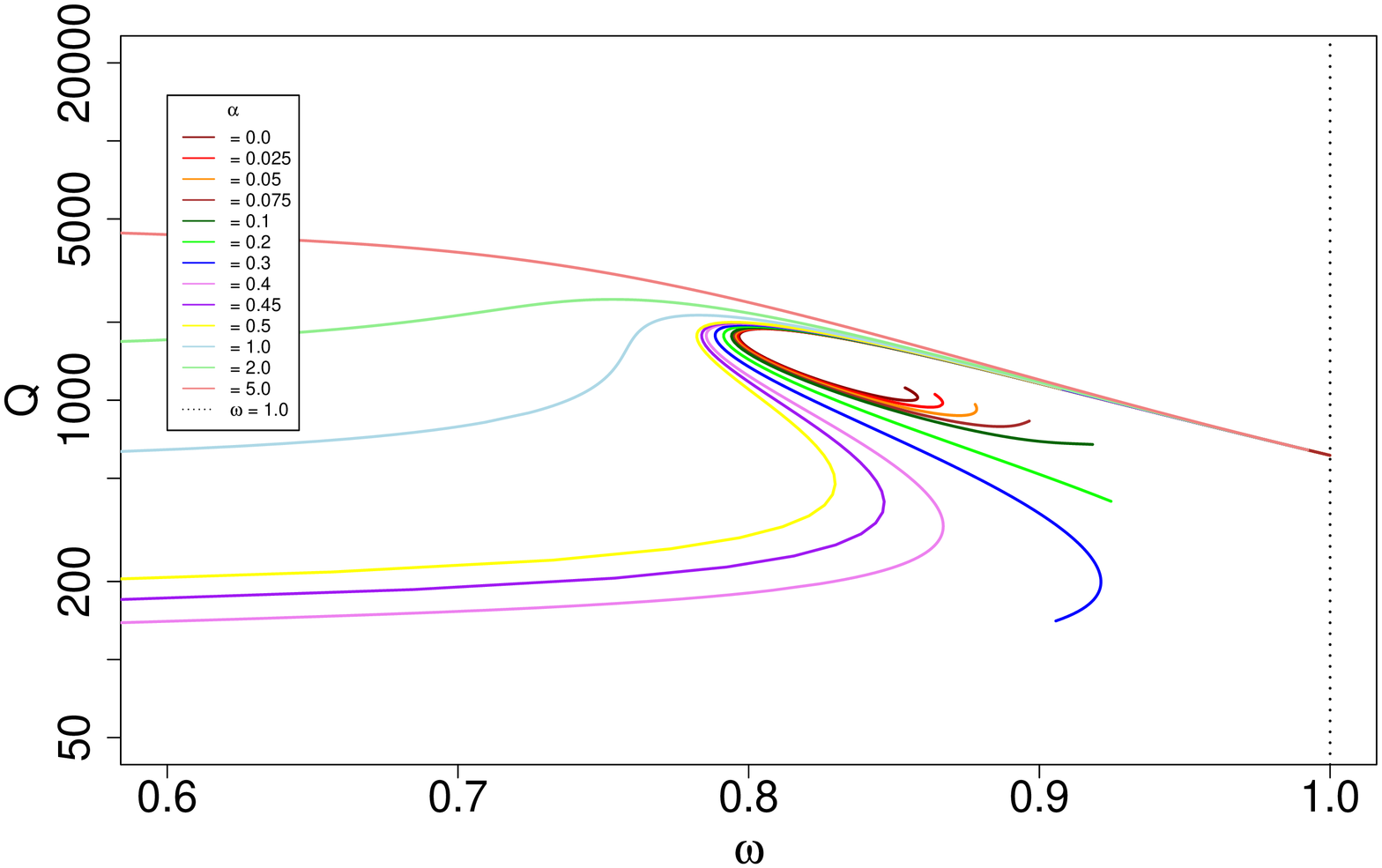}}
\end{center}
\caption{We give the charge $Q$ as function of $\omega$ for different values of $\alpha$ and
$\kappa=0.02$ (left) and $\kappa=0.05$ (right), respectively. 
\label{fig_charge}
}
\end{figure}

\begin{figure}[h!]
\begin{center}
\subfigure[\ $\kappa=0.02$]
{\label{fig_mass1}\includegraphics[width=5.5cm,angle=270]{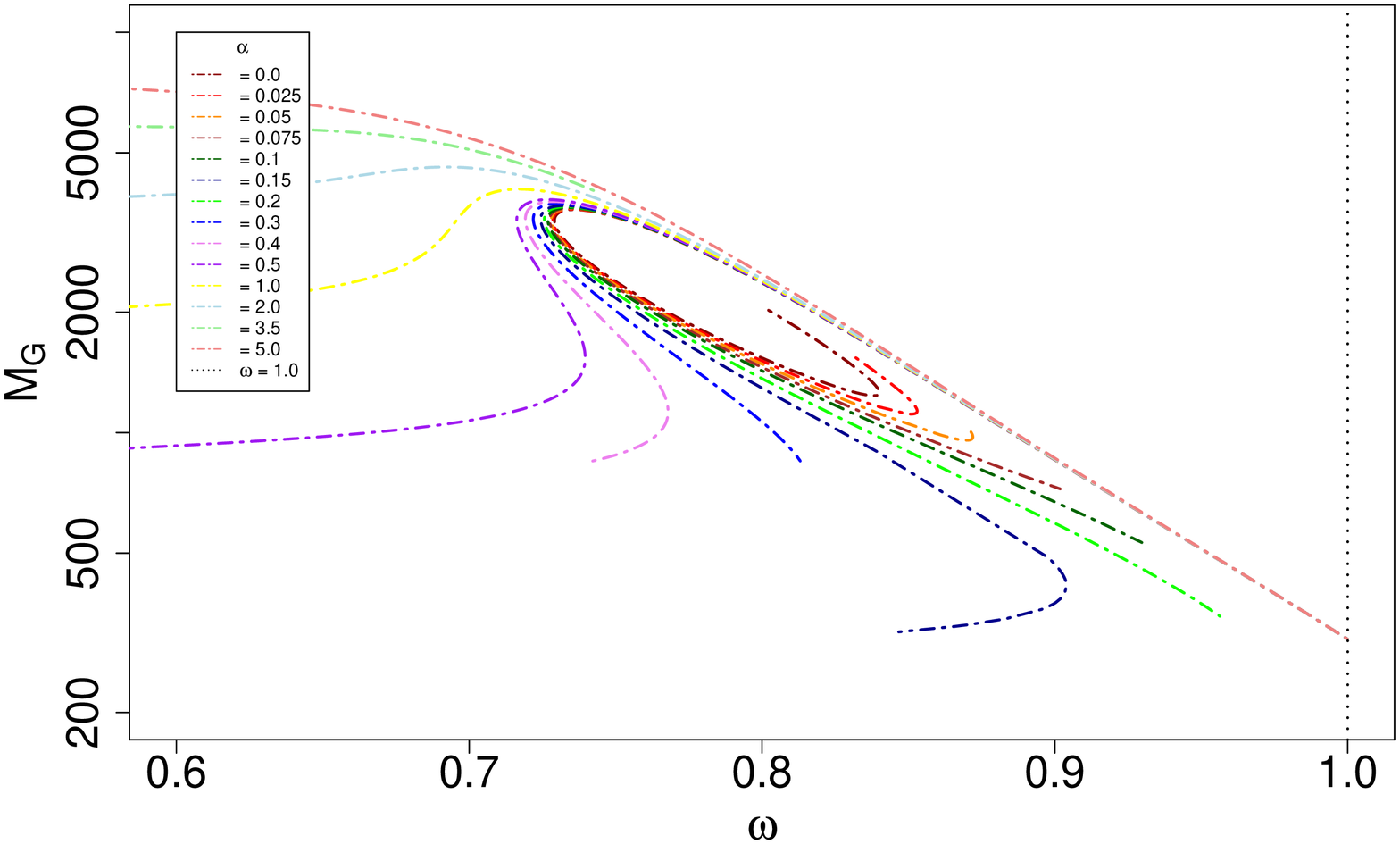}}
\subfigure[\ $\kappa=0.05$]
{\label{fig_mass2}\includegraphics[width=5.5cm,angle=270]{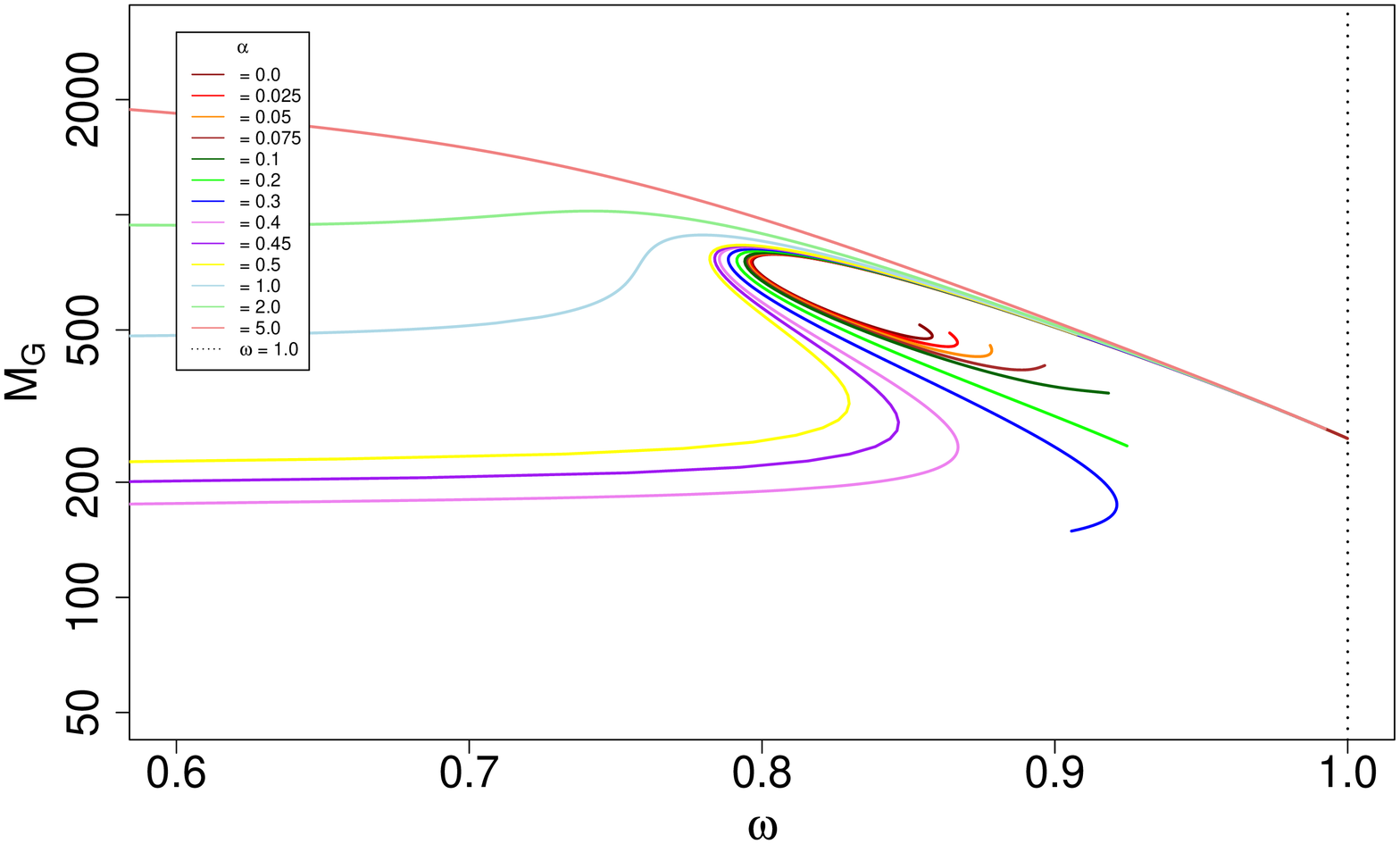}}
\end{center}
\caption{We give the gravitational mass $M_G$ as function of $\omega$ for different values of $\alpha$ and
$\kappa=0.02$ (left) and $\kappa=0.05$ (right), respectively. 
\label{fig_mass}
}
\end{figure}

\section{Numerical results}
The solutions to the coupled system of nonlinear differential equations 
are only known numerically. We have solved these equations
using the ODE solver COLSYS \cite{colsys}. 
The solutions have relative errors on the order of $10^{-6}-10^{-10}$.

In Fig.\ref{fig_charge} and Fig.\ref{fig_mass} we give the charge $Q$ and the gravitational mass $M_{\rm G}$, 
respectively,
as function of $\omega$ for two different values of $\kappa$ and
a range of values of the Gauss-Bonnet parameter $\alpha$. We observe that the maximal value $\omega_{\rm max}$
up to where the boson stars exist does not depend on the value of $\alpha$ and $\kappa$. This is not
surprising since $\omega\rightarrow \omega_{\rm max}$ is the thick wall limit with $f(r)\approx 0$ over all space. Hence, details
of the gravity model do not matter in this limit. The thin wall limit, on the other hand, is strongly influenced
by the choice of the Gauss-Bonnet coupling. This is clearly seen in Fig.\ref{fig_charge} and Fig.\ref{fig_mass}.
For small values of $\alpha$ we find that the behaviour is similar to the $\alpha=0$ limit of Einstein gravity.
The solutions exist down to a minimal value of frequency, $\omega_{\rm min}$, and from there a second branch of
solutions exists extending backwards in $\omega$ up to a critical value $\omega_{\rm cr}$ where a third branch of
solutions and consecutively a spiraling behaviour appears such that the charge $Q$ and the mass $M_{\rm G}$ are
higher on the third branch as compared to the second branch. When increasing the Gauss-Bonnet coupling $\alpha$ we find that
this spiraling behaviour disappears for $\alpha$ large enough. The actual value of $\alpha$ where this happens
is hard to determine precisely, but all our numerical results indicate that the larger $\kappa$ the larger we have
to choose $\alpha$ to see the spiraling behaviour disappear. For $\alpha$ increasing we then find that
the solutions on the second branch exist up to a critical value $\omega_{\rm cr}$ and that from there
a third branch extends backwards in $\omega$, but this time the charge and mass of the solutions on the third
branch is lower than that of the solutions on the second branch. Hence we observe that the ``spiral unfolds''.
Finally, we find that increasing $\alpha$ even further we are left with only one branch of solutions.

\begin{figure}[h!]
\begin{center}
\subfigure[\ $\kappa=0.02$]
{\label{fig_phi0_1}\includegraphics[width=5.5cm,angle=270]{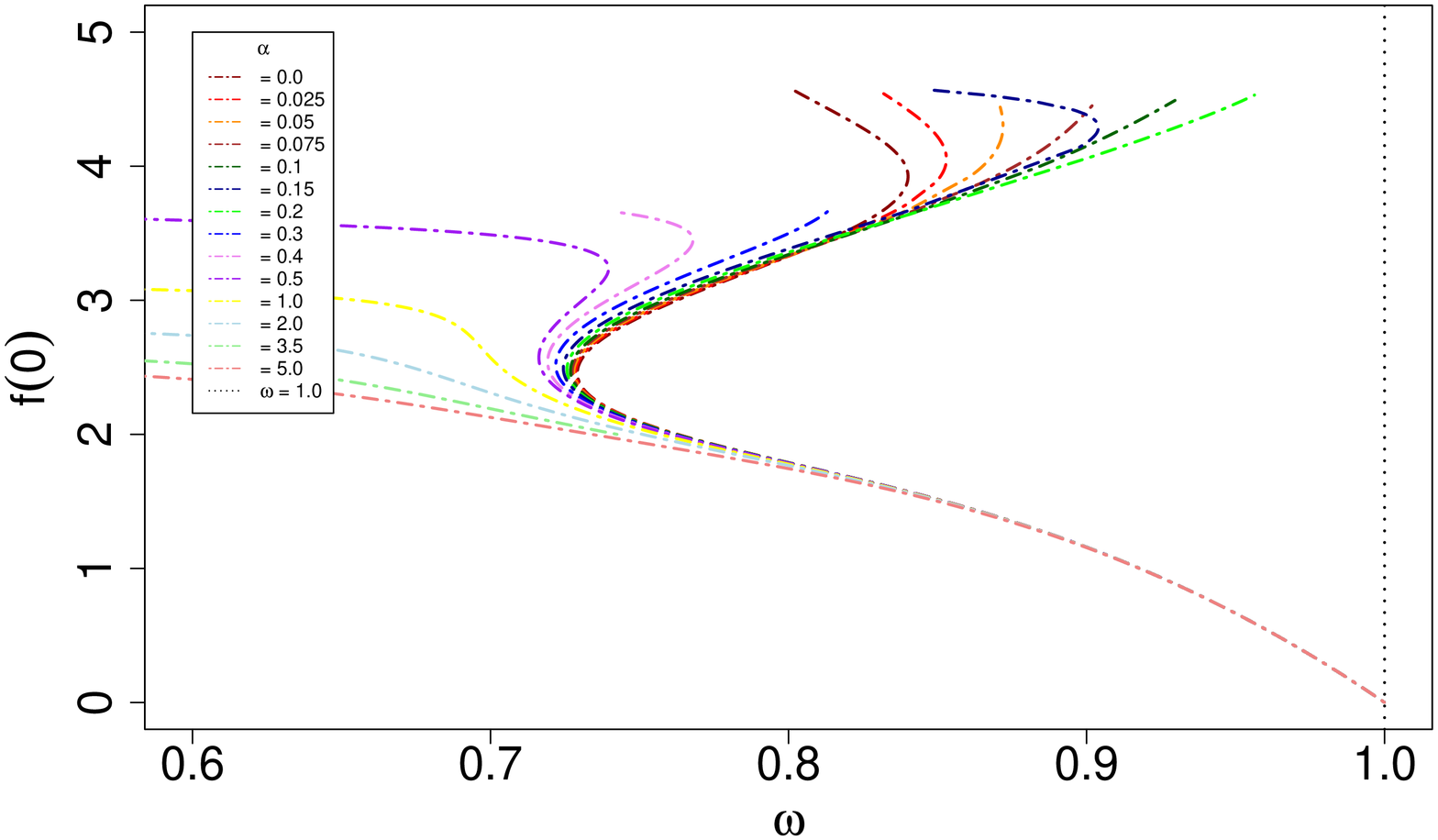}}
\subfigure[\  $\kappa=0.05$]
{\label{fig_phi0_2}\includegraphics[width=5.5cm,angle=270]{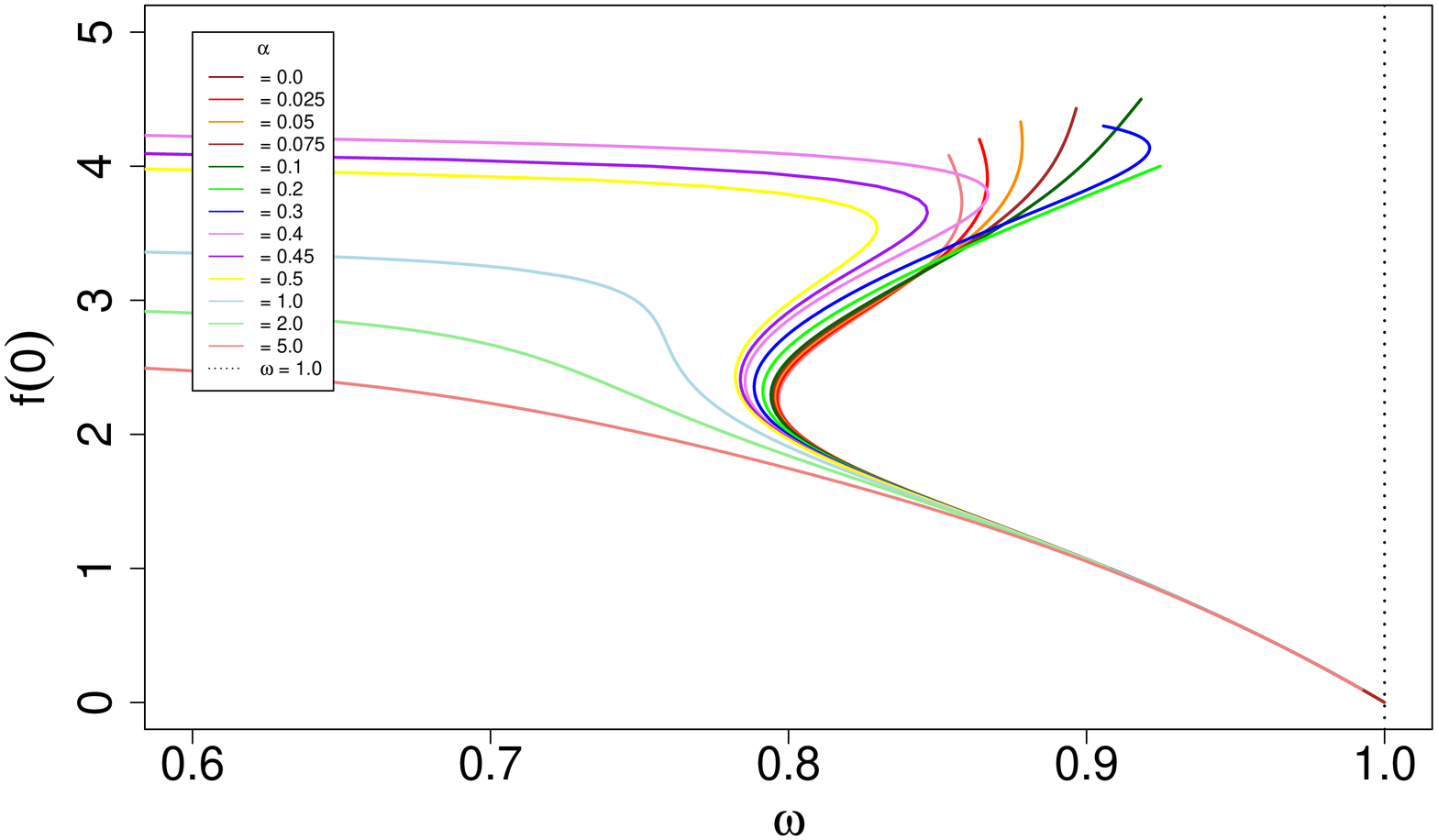}}
\end{center}
\caption{We give the central value $f(0)$ as function of $\omega$ for different values of $\alpha$ and
$\kappa=0.02$ (left) and $\kappa=0.05$ (right), respectively. 
\label{fig_phi0}
}
\end{figure}

\begin{figure}[h!]
\begin{center}
\includegraphics[width=7cm,angle=270]{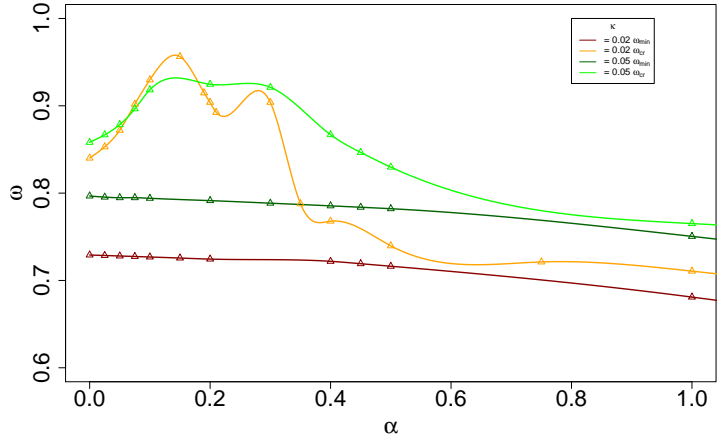}
\end{center}
\caption{\label{boson} We show $\omega_{min}$ as well as $\omega_{cr}$ (which corresponds to
the value of $\omega$ where the second and third branch of solutions join) as function of $\alpha$ 
for two different values of $\kappa$. }
\end{figure}

To understand this pattern in more detail, we have also plotted the value of the scalar field function
at the origin, $f(0)$, as function of $\omega$. Our results for several values of $\alpha$ are given in Fig.\ref{fig_phi0}.
For small $\alpha$, we find the spiraling behaviour characteristic for boson stars in Einstein gravity with
$\phi(0)\rightarrow \infty$ at the center of the spiral in Fig.\ref{fig_charge} or Fig.\ref{fig_mass}. This
latter case corresponds to the thin wall limit with $f(r)$ strongly peaked at $r=0$. 
For large values of $\alpha$ we find that the range of values of $f(0)$ is limited. While $\omega$ tends to
smaller and smaller values $f(0)$ stays nearly constant. The larger $\alpha$ the smaller is this nearly constant
value of $f(0)$. In other words: as soon as the Gauss-Bonnet coupling is large enough we cannot construct ``thin-wall'' boson
stars. 

Let us remark that the numerical calculations become very tedious at the turning points of the curves
and it is difficult to determine the exact values of $\omega_{\rm min}$ and $\omega_{\rm cr}$. However,
we have plotted our existing results in order to understand the qualitative pattern of the solutions.
In Fig.\ref{boson} we give the value of $\omega_{\rm min}$ as well as $\omega_{\rm cr}$, which corresponds to
the value of $\omega$ where the second and third branch join, as function of $\alpha$ for two different
values of $\kappa$. We find that $\omega_{\rm min}$ decreases continuously with $\alpha$, 
while $\omega_{\rm cr}$ shows a more complicated behaviour. 
First the value of $\omega_{\rm cr}$ increases, but can of course not increase without bound
since $\omega=1$ is the limiting value for
$\omega$ up to where solutions can exist. Then $\omega_{\rm cr}$ shows an oscillating behaviour
for intermediate values of $\alpha$ and finally decreases. For large enough $\alpha$ 
we find that $\omega_{\rm cr}$ becomes equal to $\omega_{\rm min}$ such that only one branch of solutions exists.

\begin{figure}[h!]
\begin{center}
\subfigure[\ $\kappa=0.02$]
{\label{fig_phi0_1}\includegraphics[width=5.5cm,angle=270]{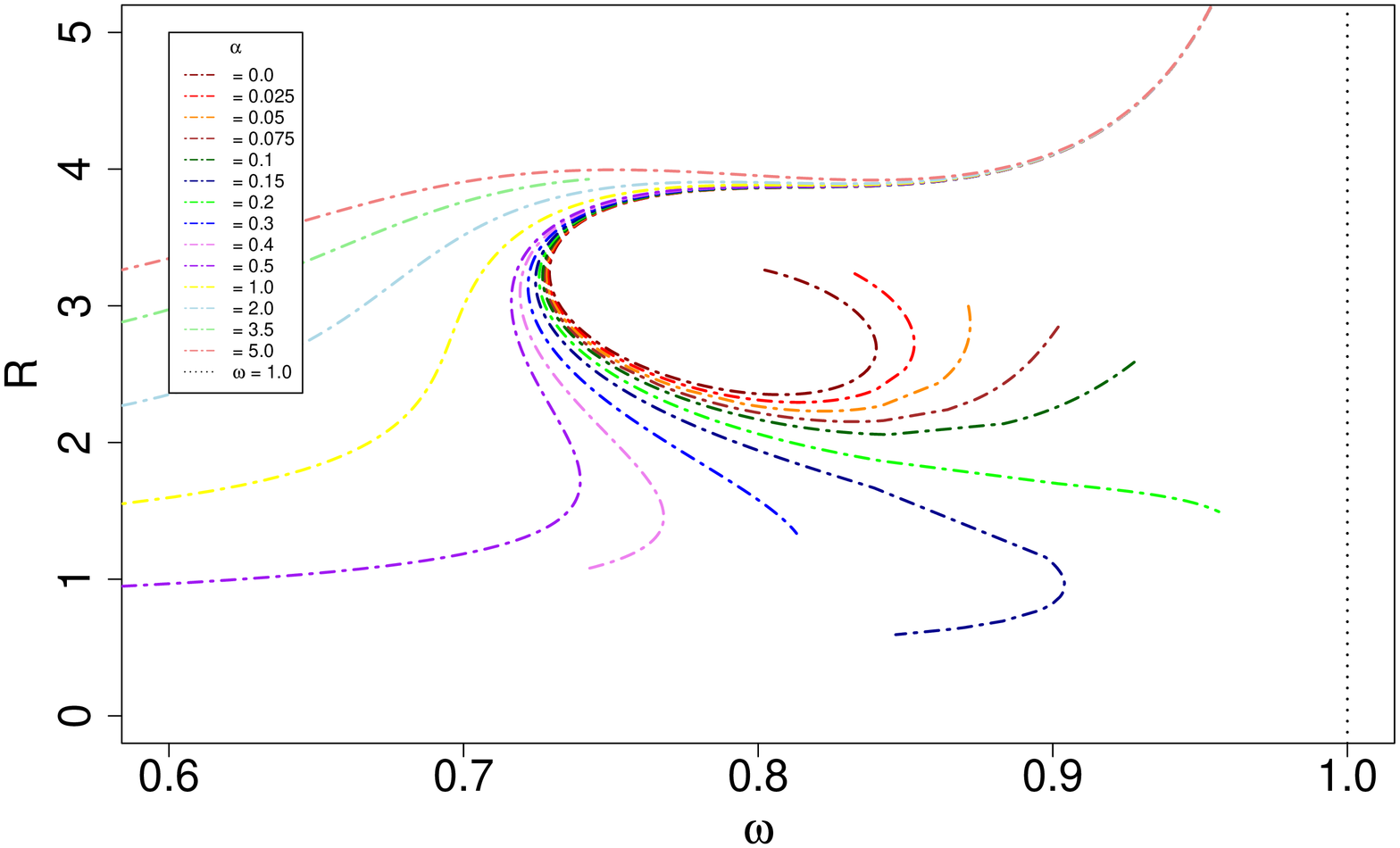}}
\subfigure[\  $\kappa=0.05$]
{\label{fig_phi0_2}\includegraphics[width=5.5cm,angle=270]{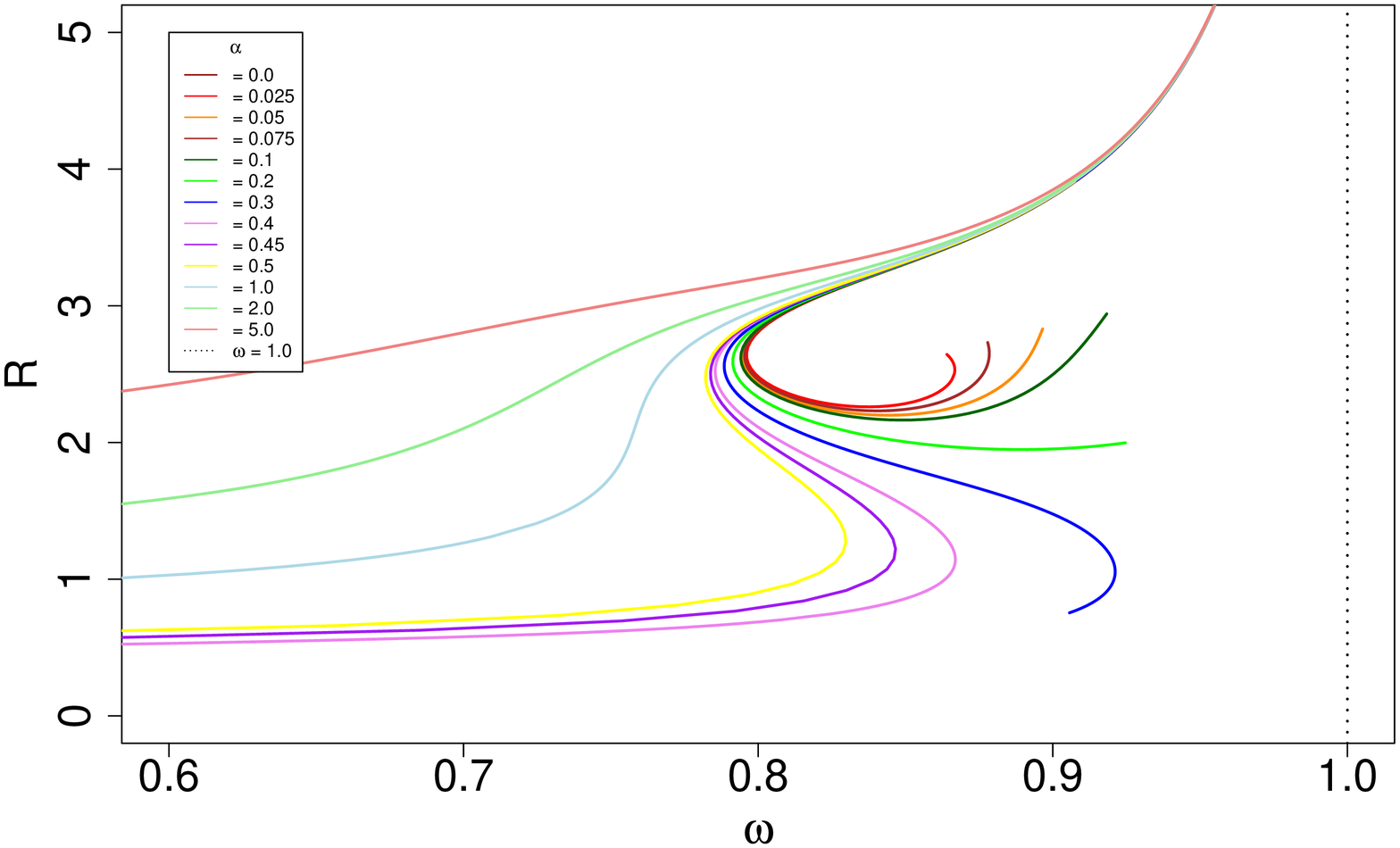}}
\end{center}
\caption{\label{radius_omega} We give the radius $R$ as function of the frequency $\omega$ for different values of
$\alpha$ and $\kappa=0.02$ (left) and $\kappa=0.05$ (right), respectively.  }
\end{figure}

In order to have an idea of the sizes of Gauss-Bonnet boson stars in comparison to their Einstein counterparts,
we have also computed the radius of these objects using (\ref{radius}). In Fig.\ref{radius_omega} we shown the
value of $R$ in dependence on $\omega$ for different values of $\alpha$ and two values of $\kappa$. 
We find that the range of radii possible first increases with increasing $\alpha$ with 
the largest range of possible $R$ at $\alpha\approx 0.3$, but that then the range of possible $R$ is narrowing again until it becomes smaller than in the
Einstein gravity case for $\alpha \gtrsim 2$. 

\begin{figure}[h!]
\begin{center}
\subfigure[\ $\kappa=0.02$]
{\label{radius_mass_1}\includegraphics[width=5.5cm,angle=270]{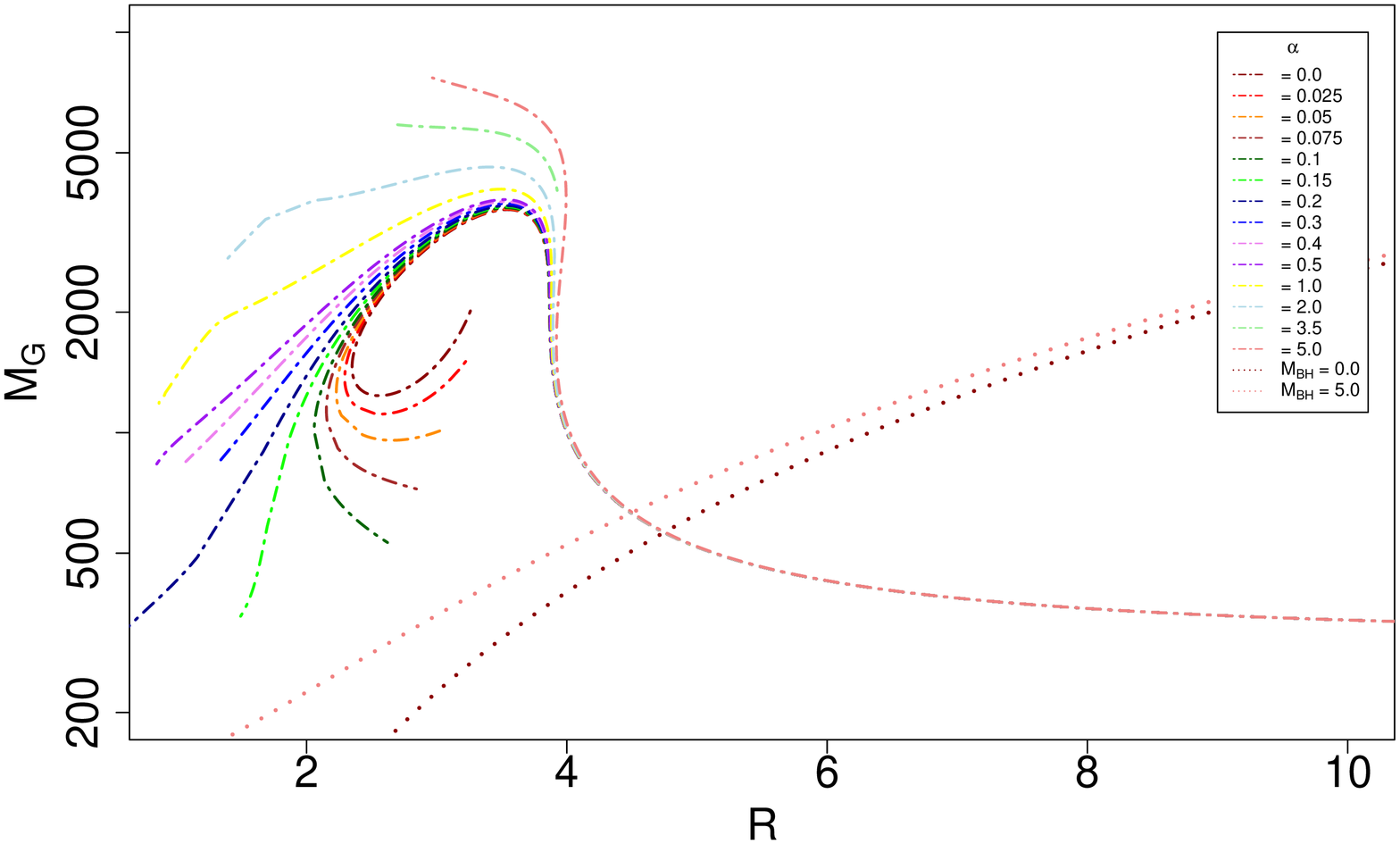}}
\subfigure[\  $\kappa=0.05$]
{\label{radius_mass_2}\includegraphics[width=5.5cm,angle=270]{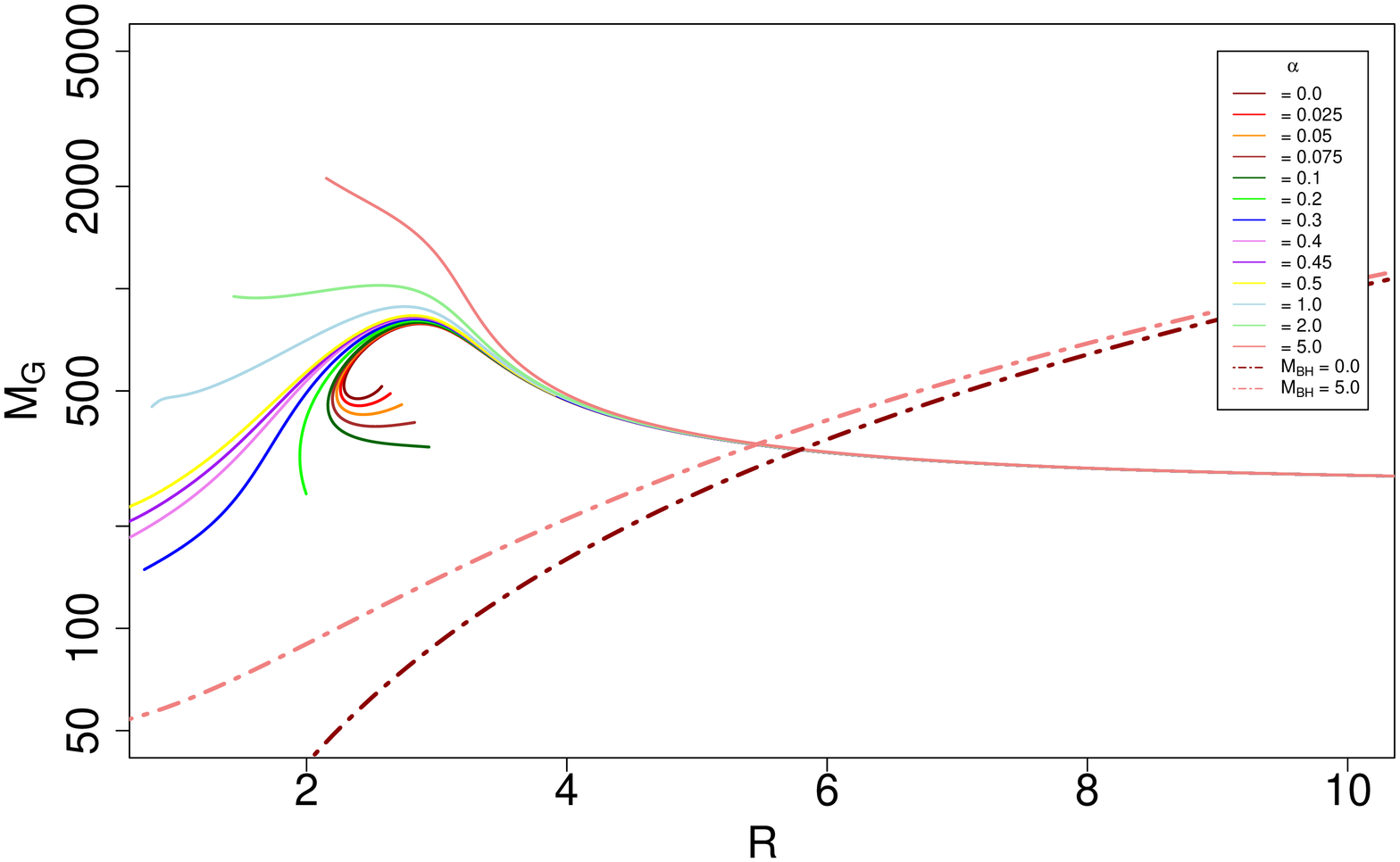}}
\end{center}
\caption{\label{radius_mass} 
We give the gravitational mass $M_G$ as function of the radius $R$ for different values of
$\alpha$ and $\kappa=0.02$ (a) and $\kappa=0.05$ (b), respectively. 
We compare this with the radius-mass relation for the corresponding black hole solutions existing for $f\equiv 0$.
In this latter case, the gravitational mass of the black hole is given in terms of the radius of the event horizon $r_h$ as follows 
$M_G=M/\kappa=(r_h^2+ \alpha)/(2\kappa)$. }
\end{figure}

\subsection{Physical values and comparison to neutron stars}
Here we would like to briefly comment on the physical quantities of boson stars and the relation to neutron stars 
which we stated above might
be well ``toy-modeled by the former. First of all, let us make a connection between the dimensionless quantities
used in our paper and the physical values (indicated by a subscript ``phys'' in the following). The mass
and radius are
\begin{equation}
 \label{physical_mass}
M_{\rm phys}=10^{27} {\rm kg} \left(\frac{6.67\cdot 10^{-11} {\rm Nm^3/kg}}{G_5}\right)
\left(\frac{m}{10^{-7} {\rm eV}}\right) M_{\rm G} \cdot \kappa  \ ,
\end{equation}
and
\begin{equation}
 \label{physical_radius}
R_{\rm phys} = 10^{-3} {\rm km} \left(\frac{10^{-7} {\rm eV}}{m}\right) \cdot R  \ ,
\end{equation}
where $m$ is the mass of the scalar field in {\rm eV} and we have taken into account that Newton's constant
$G_5$ in (4+1) dimensions might have a different value than the one in (3+1) dimensions. As indicated in Fig.\ref{radius_mass}
we find that the Gauss-Bonnet boson stars are very dense and exceed e.g. the density of black holes. In some
sense this is not surprising, because our boson stars are non-compact objects without
a definite radius beyond which the energy density and pressure become strictly zero. Hence different definitions
of the radius can be considered. However, it is interesting that the mass-radius relation becomes {\it unique}
for a sufficiently high Gauss-Bonnet coupling in the sense that a given radius uniquely determines the mass of the
boson star. This is different in the Einstein-Hilbert case, where a spiraling behaviour leads to several boson star solutions
with different masses existing for one given value of the radius. Neutron stars possess (in most cases) 
a unique relation between the mass and the radius
that has to be determined using the equation of state (see e.g. \cite{Lattimer:2004pg,Lattimer:2006xb} 
and references therein). Interestingly, the qualitative features of the mass-radius relation
of neutron stars seem to resemble some of the mass-radius relations we obtain for large enough values of
$\alpha$ (compare e.g. Fig. \ref{radius_mass} in this paper and Fig.2 in \cite{Lattimer:2004pg}).
In \cite{Pani:2011xm} perfect fluid stars in (3+1)-dimensional Einstein-Gauss-Bonnet-dilaton theory
were considered. The mass-radius diagram looks similar to what we find.

\section{Summary and Outlook}
We have constructed Gauss-Bonnet boson stars in (4+1)-dimensional space-time and have investigated their properties
depending on the coupling constants in the model. Since we were aiming at studying the sole effect of the
Gauss-Bonnet term without taking into account a dilaton (4+1) dimensions is the lowest number of dimensions
where this higher gravity term has an effect. We find that a qualitative change happens when the
Gauss-Bonnet parameter $\alpha$ is large enough. While in Einstein-Hilbert gravity the solutions reach a ``thin wall''
limit with the scalar field at the center of the star diverging, we cannot reach a ``thin wall'' limit
for $\alpha$ large enough. Furthermore, the spiraling behaviour characteristic for boson stars in Einstein
gravity disappears for $\alpha$ large enough and a unique mass-radius relation for the Gauss-Bonnet boson stars 
is found. Surely, if we want to compare this to astrophysical objects and promote the idea that boson stars
could act as toy models for neutron stars, we would have to study these effects in (3+1) dimensions. However, this
would require an additional scalar field, the dilaton. It would then be interesting to see whether the
qualitative features of the mass-radius relation that resemble those of ``real'' neutron stars will still be
present. In \cite{Pani:2011xm} this was investigated for perfect fluid stars and constraints on the Gauss-Bonnet
parameter were obtained. 

It would also be interesting to extend our results to asymptotically Anti-de Sitter (AdS) space-time.
In the context of the AdS/CFT correspondence  \cite{ggdual,adscft} planar boson stars in AdS
have been interpreted
as Bose-Einstein condensates of glueballs \cite{horowitz,hartmann_riedel1}. However, these solutions
have only been considered in Einstein gravity which corresponds to the large $N$ limit on the Quantum Field
Theory side. Hence it would be interesting to take higher order curvature corrections into account.
Our model would than holographically describe a strongly coupled Quantum Field theory away from $N=\infty$ on a 3-sphere. 
This is currently under investigation. 

Finally, it would be interesting to study the rotating counterparts of our solutions. In (3+1) dimensions,
rotating boson stars are necessarily axially symmetric and partial differential equations have to be solved.
In (4+1) dimensions two orthogonal planes of rotation exist. If the two angular momenta associated to these planes
would be chosen equal, the symmetry of the system can be enhanced and ``only'' ordinary differential equations
need to be solved \cite{hartmann_kleihaus_kunz_list}. 

\vspace{0.5cm}
{\bf Acknowledgment}

\noindent
We gratefully acknowledge support within the DFG Research
Training Group 1620 {\it Models of Gravity}.

\end{document}